\documentclass[preprint, 3p]{elsarticle}
\journal{PAMM}

\usepackage[T1]{fontenc}
\usepackage[utf8]{inputenc}
\usepackage[acronym]{glossaries}
\usepackage{bm}
\usepackage{nicefrac}
\usepackage{enumitem}
\usepackage{amsmath}
\usepackage{mathtools}
\usepackage{amsfonts}
\usepackage{siunitx}
\usepackage{ulem}
\usepackage{orcidlink}
\usepackage{cleveref}

\newcommand*{\tran}{^{\mkern-1.5mu\mathsf{T}}} 
\newacronym{uav}{UAV}{Unmanned Aerial Vehicle}
\newacronym{mpc}{MPC}{Model Predictive Control}
\newacronym{mppfc}{MPPFC}{Model Predictive Path-Following Control}
\newacronym{ocp}{OCP}{Optimal Control Problem}
\newacronym{nlp}{NLP}{nonlinear program}
\newacronym{pid}{PID}{Proportional-Integral-Derivative}
\newacronym{pwm}{PWM}{Pulse-Width-Modulation}
\newacronym{igc}{IGC}{Integrated Guidance and Control}
\newacronym{sgc}{SGC}{Separated Guidance and Control}
\newacronym{imu}{IMU}{Inertial Measurement Unit}
\newacronym{sitl}{SITL}{Software-in-the-Loop}
\newacronym{ros2}{ROS 2}{Robot Operating System 2}
\newacronym{rk4}{RK4}{fourth-order Runge-Kutta method}
\newacronym{fatrop}{FATROP}{Fast Trajectory Optimizer}
\newacronym{ipopt}{IPOPT}{Interior Point Optimizer}

\begin{document}
\begin{frontmatter}
\title{Model Predictive Path-Following Control for a Quadrotor} 

\author{David Leprich\,\orcidlink{0009-0007-7080-9453}} 
\author{Mario Rosenfelder\,\orcidlink{0000-0003-0460-0612}} 
\author{Mario Hermle\,\orcidlink{0009-0004-7937-1943}} 
\author{Jingshan Chen\,\orcidlink{0000-0002-5617-0469}} 
\author{Peter Eberhard\,\orcidlink{0000-0003-1809-4407}} 

\address{Institute of Engineering and Computational Mechanics~(ITM), University of Stuttgart, Germany,\\
	(e-mail: \lbrack david.leprich, mario.rosenfelder, mario.hermle, jingshan.chen, peter.eberhard\rbrack @itm.uni-stuttgart.de)}

\begin{abstract}
Automating drone-assisted processes is a complex task. 
Many solutions rely on trajectory generation and tracking, whereas in contrast, path-following control is a particularly promising approach, offering an intuitive and natural approach to automate tasks for drones and other vehicles.
While different solutions to the path-following problem have been proposed, most of them lack the capability to explicitly handle state and input constraints, are formulated in a conservative two-stage approach, or are only applicable to linear systems.
To address these challenges, the paper is built upon a \acrlong{mpc}-based path-following framework and extends its application to the Crazyflie quadrotor, which is investigated in hardware experiments.
A cascaded control structure including an underlying attitude controller is included in the \acrlong{mppfc} formulation to meet the challenging real-time demands of quadrotor control.
The effectiveness of the proposed method is demonstrated through real-world experiments, representing, to the best of the authors' knowledge, a novel application of this MPC-based path-following approach to the quadrotor.
Additionally, as an extension to the original method, to allow for deviations of the path in cases where the precise following of the path might be overly restrictive, a corridor path-following approach is presented.
\end{abstract}


\end{frontmatter}

\section{Introduction}\label{sec:Introduction}
In recent years, there has been a growing interest in quadrotors from both academia and industry.
Their inherent adaptability and versatility make them ideal candidates for a wide range of applications within the domain of \acrlong*{uav}s~(\acrshort*{uav}s). 
These applications include, but are not limited to, delivery services, infrastructure inspection, precision agriculture, search and rescue operations, as well as border patrol and surveillance activities.
A crucial aspect of automating such processes is the capability to autonomously follow a problem-specific state reference.
In general, it is important to distinguish between references provided as a trajectory and those provided as a path.
The former is a time-dependent signal, providing state references for each time instant, while the latter is a geometric curve, defined completely independent of time.
In many applications, the latter is a more intuitive approach for providing state references, since it  allows for a decoupling of spatial information from the complexity of assigning precise timing to each state reference.
An illustrative example is the autonomous navigation through agricultural fields to detect pests and implement measures for their control and management~\cite{MaslekarKulkarniChakravarthy20}.
In such a scenario, the flight plan might be defined as a geometric path, without having to consider any time information.
This is attributed to the fact that the precise timing of the quadrotor's motion is there of secondary importance compared to the accurate tracking of the reference path.
The path is subsequently converted into a time-dependent trajectory through the application of a timing law, which dictates the temporal evolution of the path parameter, thereby rendering it time-dependent.
While designing a geometric path is an intuitive approach to provide motion specifications, finding a suitable timing law that satisfies both hardware constraints and dynamic feasibility, while not being excessively conservative, often necessitates the use of advanced algorithms~\cite{SkjetneFossenKokotovic04}.
Determining the appropriate timing law is typically done as a pre-processing step prior to addressing the actual control problem, as shown in~\cite{ChenXuEbelEtAl25}.
There, the authors propose a path-following controller for a quadrotor in three-dimensional space.
In a pre-processing step, the timing law is designed using a differential equation which guarantees the satisfaction of dynamical constraints and makes the time derivative of the path parameter converge to a desired value.
Rendering the path-parameter time-dependent, the resulting trajectory is then tracked by the quadrotor using a backstepping controller.
This two-fold approach might be conservative and non-optimal, as the timing law is designed independently of the controller, especially if the dynamics of the system are not considered explicitly.
Note that in other works, one-stage and two-stage approaches refer to the process of trajectory generation \cite{ChenXuEbelEtAl25}, while this work refers to these as the complete process of trajectory generation and tracking.
Other approaches include learning-based methods such as deep learning techniques~\cite{RubiMorcegoPerez20} and geometric techniques like Carrot-Chasing~\cite{PerezLeonAcevedoMillanRomeraEtAl20}, to only name a few.
An extensive overview of path-following control techniques for quadrotors can be found in \cite{RubiPerezMorcego20}.

Most of the aforementioned methods lack the consideration of state and input constraints as well as of the dynamics of the system and employ a rather conservative two-stage approach.
In contrast to the two-stage approach, in \cite{RozaMaggiore12} a one-stage approach is presented, which does not employ a timing law but rather uses a feedback linearization technique to turn the zero dynamics manifold of the system into the path following manifold and render it asymptotically stable.
While this one-stage approach represents a less conservative solution, it does not consider the system's state and input constraints.
Therefore, this work focuses on overcoming these limitations, by novelly applying a one-stage \acrfull*{mpc} based path-following approach\cite{Faulwasser12} to a quadrotor.
Besides the explicit consideration of state and input constraints, additional advantages of an \acrshort*{mpc}-based path-following approach include theoretical stability guarantees for asymptotical convergence to the end of the path, as well as a straightforward user-defined design and tuning of the control scheme.
Next to the asymptotical convergence, multiple examples in~\cite{FaulwasserFindeisen16, FaulwasserWeberZometaEtAl17} exhibit rapid transient performance of the proposed \acrshort{mpc} scheme, even when the system initially starts besides the path.
Its successful application to planar robots~\cite{FaulwasserFindeisen16} and industrial robots~\cite{FaulwasserWeberZometaEtAl17} elicits interest in its application to quadrotors.
To the best knowledge of the authors, this is the first time that the path-following \acrshort*{mpc} scheme in \cite{Faulwasser12} is applied to a quadrotor in real-world experiments, in particular here to the popular Crazyflie quadrotor.
Furthermore, the scheme is also extended to the less conservative case of corridor path-following and validated experimentally.
A similar approach referred to as Model Predictive Contouring Control is presented in \cite{RomeroSunFoehnEtAl22} to achieve time-optimal tracking of a path defined by a set of waypoints, with real-world experimental validation.
Related to this work are \cite{WangPanHuEtAl19, WangZhaoHuEtAl20, SanchezDJorgeRaffoEtAl21}, where a different \acrshort*{mpc} scheme is applied, to solve the path-following problem for a quadrotor in simulation only.
Additionally, in \cite{WeiZhengLiEtAl24, YangZhengPanEtAl21} a learning-based \acrshort{mpc} approach is provided, with real-world experimental validation in \cite{WeiZhengLiEtAl24}. 

The paper is organized as follows. 
In~\Cref{sec:ModellingAndControl} the equations of motion for a Crazyflie quadrotor are derived. 
Additionally, the employed controller structure is explained.
Subsequently, in~\Cref{sec:PathFollowing}, the path-following problem is defined and the corresponding \acrshort{mpc} scheme is presented.
Then, in~\Cref{sec:ExperimentalResults}, the experimental setup is described and the results of the conducted experiments are presented.
\Cref{sec:Conclusion} summarizes the findings and provides an outlook to future work.

\section{Modelling and Attitude Control of a Crazyflie Quadrotor}\label{sec:ModellingAndControl}
\begin{figure}[ht]
    \centering
    \includegraphics{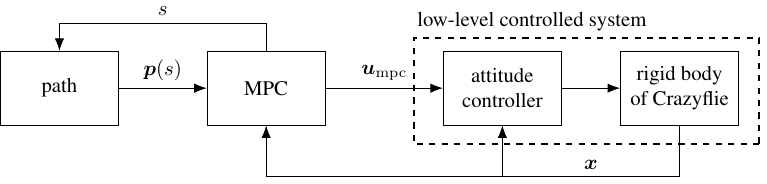}
    \caption{Complete two-layered control structure.}
    \label{fig:CompleteControlLoop}
\end{figure}

Most of the control approaches for quadrotors can be categorized into two groups, \acrfull{igc} and \acrfull*{sgc}~\cite{RubiPerezMorcego20}.
The \acrshort{igc} approach focuses on the design of a single control loop, where the controller provides \acrfull{pwm} signals to the motors of the quadrotor.
Due to the unstable dynamics of the quadrotor and aerodynamic disturbances, such a control structure needs to run at very high frequencies to ensure robust stability and good disturbance rejection.
The processing and computation of motor signals at high frequencies present a significant challenge for optimization-based control frameworks, such as \acrshort{mpc}, particularly in the presence of nonlinear system dynamics \cite{GrosZanonQuirynenEtAl20}.
Therefore, such an approach to the path-following problem might be computationally infeasible.
As a consequence, this work focuses on utilizing the \acrfull*{sgc} approach, where the control structure is broken down into a two-layered cascaded control architecture.
The outer control loop utilizes a sophisticated \acrshort*{mpc} scheme to provide guidances signals to an inner control loop tasked with stabilizing the attitude of the Crazyflie quadrotor.
The complete control structure is depicted in \Cref{fig:CompleteControlLoop}.

\noindent
To obtain the equations of motion for the Crazyflie, it is essential to examine both the translational and rotational dynamics.
Considering only the main body of the quadrotor, see Figure~\ref{fig:CrazyflieCoordinateSystem}, it has a total of six degrees of freedom, three corresponding to translational motion and three to rotational motion.
In a first step, the kinematics of the quadrotor is derived.
Subsequently, an in-depth explanation of the onboard attitude control of the low-level controlled system is provided. 
This is a necessary intermediate step to be able to derive the equations of motion for the quadrotor, since the rotational dynamics of the Crazyflie is heavily influenced by the usage of the attitude controller.
Lastly, the translational and rotational dynamics are derived and transformed into a state-space representation.

\subsection{Kinematics}
\begin{figure}[ht]
    \centering
    \includegraphics{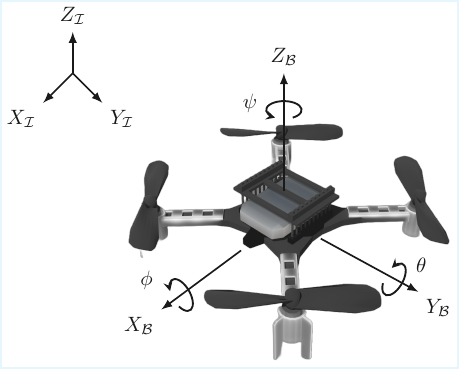}
    \includegraphics[scale=0.2]{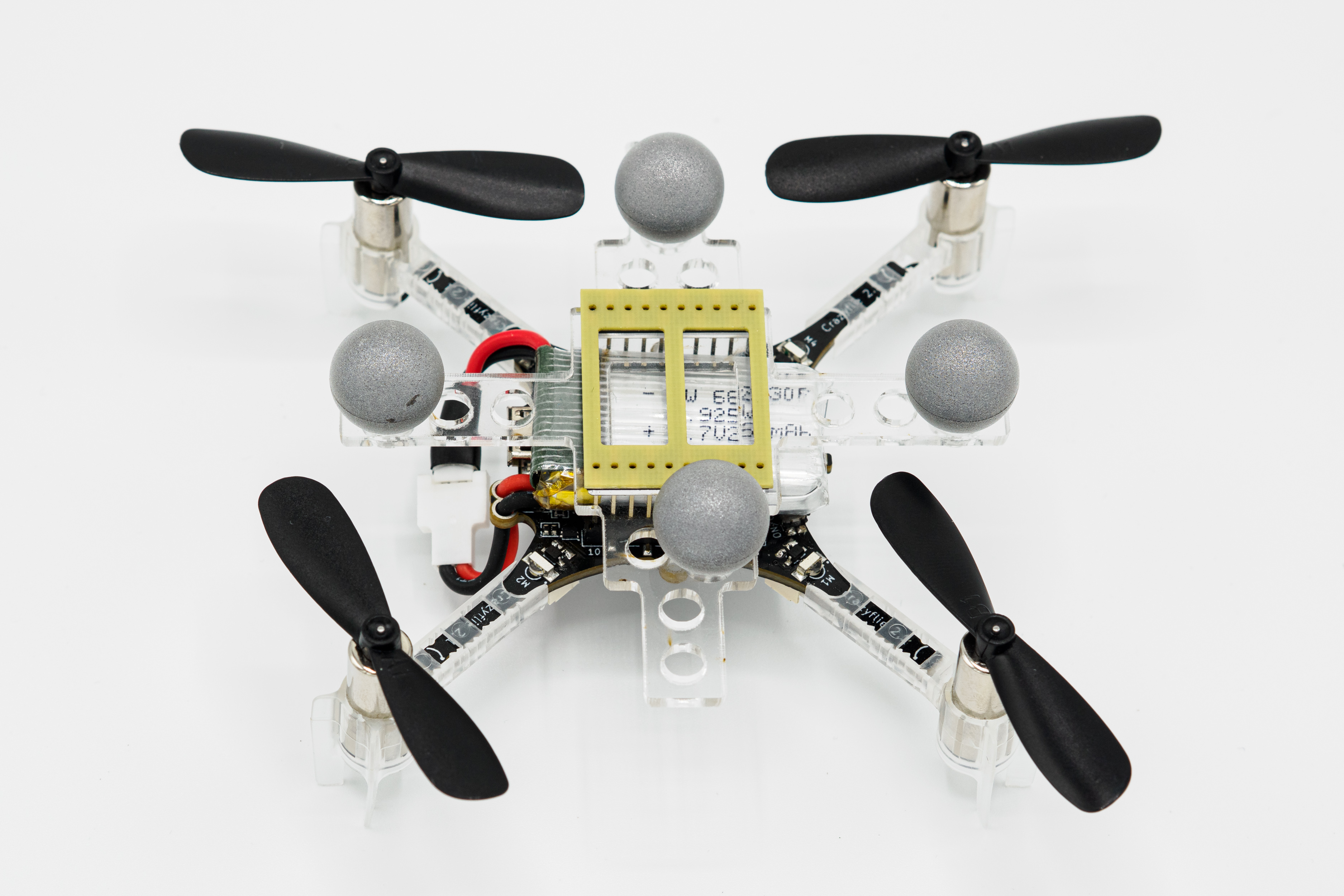}
    \caption{Crazyflie quadrotor with the body fixed coordinate system $\mathcal{B}$ and the global reference frame $\mathcal{I}$, denoted by $\left(\cdot\right)_\mathcal{B}$ and~$\left(\cdot\right)_\mathcal{I}$, respectively.}
    \label{fig:CrazyflieCoordinateSystem}
\end{figure}

The derivation of the nonlinear equations of motion involves the introduction of two coordinate systems. 
The coordinate system $\mathcal{I}$ denotes the global reference coordinate system that is fixed in space, while the coordinate system $\mathcal{B}$ is attached to the quadrotor's body, with its origin positioned at the centre of mass and its $X_\mathcal{B}$-axis pointing along the forward direction. 
An illustration depicting these coordinate systems is shown in~\Cref{fig:CrazyflieCoordinateSystem}.
The displacement of the origin of $\mathcal{B}$ with respect to the origin of $\mathcal{I}$ and the translational velocity are represented by 
\begin{equation}
    \bm{\xi} = \begin{bmatrix}
        x & y & z
    \end{bmatrix}\tran
	\qquad\text{and}\qquad    
    \dot{\bm{\xi}} = \begin{bmatrix}
        \dot{x} & \dot{y} & \dot{z}
    \end{bmatrix}\tran ,
\end{equation}
respectively. Additionally, the three rotational degrees of freedom are described by successive rotations around each axis of $\mathcal{B}$.
Here, $\phi$ represents the roll angle around the $X_\mathcal{B}$-axis, $\theta$ represents the pitch angle around the $Y_\mathcal{B}$-axis, and $\psi$ the yaw angle around the $Z_\mathcal{B}$-axis of $\mathcal{B}$.
In this context, the vector representing the introduced angles is defined as
\begin{equation}
    \bm{\eta} = \begin{bmatrix}
        \phi & \theta & \psi
    \end{bmatrix}\tran,
\end{equation}
and will be referred to as the attitude of the quadrotor.
To describe the rotation of $\mathcal{B}$ in relation to $\mathcal{I}$, three consecutive rotations based on the Euler angle convention are applied.
The rotation matrix is therefore given by
\begin{equation}
        \bm{R}_{\mathcal{I}\mathcal{B}}(\bm{\eta}) = \bm{R}_{\mathcal{I}\mathcal{I}'}(\psi) \, \bm{R}_{\mathcal{I}^{'}\mathcal{I}^{''}}(\theta) \,\bm{R}_{\mathcal{I}^{''}\mathcal{B}}(\phi) = \begin{bmatrix}
            \mathrm{c}_\psi \mathrm{c}_\theta & \mathrm{c}_\psi \mathrm{s}_\phi \mathrm{s}_\theta - \mathrm{c}_\phi \mathrm{s}_\psi & \mathrm{s}_\phi \mathrm{s}_\psi + \mathrm{c}_\phi \mathrm{c}_\psi \mathrm{s}_\theta \\
            \mathrm{c}_\theta \mathrm{s}_\psi & \mathrm{c}_\phi \mathrm{c}_\psi + \mathrm{s}_\phi \mathrm{s}_\psi \mathrm{s}_\theta & \mathrm{c}_\phi \mathrm{s}_\psi \mathrm{s}_\theta - \mathrm{c}_\psi \mathrm{s}_\phi \\
            -\mathrm{s}_\theta & \mathrm{c}_\theta \mathrm{s}_\phi & \mathrm{c}_\phi \mathrm{c}_\theta
        \end{bmatrix},
\end{equation}
where $\mathrm{s}_x$ and $\mathrm{c}_x$ are abbreviations for $\sin{x}$ and $\cos{x}$, respectively.
Based on the elementary rotation matrices, the angular velocity of the quadrotor in the global reference frame $\mathcal{I}$ can be expressed as
\begin{equation}
    \bm{\omega}_\mathcal{I} = \begin{bmatrix}
        0 \\
        0 \\
        \dot{\psi}
    \end{bmatrix} + \bm{R}_{\mathcal{I}\mathcal{I}'}(\psi)
    \begin{bmatrix}
        0 \\
        \dot{\theta} \\
        0
    \end{bmatrix} + \bm{R}_{\mathcal{I}\mathcal{I}'}(\psi) \, \bm{R}_{\mathcal{I}^{'}\mathcal{I}^{''}}(\theta)
    \begin{bmatrix}
        \dot{\phi} \\
        0 \\
        0
    \end{bmatrix} = \underbrace{\begin{bmatrix}
        \cos{\psi}\cos{\theta} & -\sin{\theta} & 0 \\
        \sin{\psi}\cos{\theta} & \cos{\psi} & 0 \\
        -\sin{\theta} & 0 & 1
    \end{bmatrix}}_{\displaystyle \eqqcolon \bm{J}_\mathrm{R}}
    \underbrace{\begin{bmatrix}
        \dot{\phi} \\
        \dot{\theta} \\
        \dot{\psi}
    \end{bmatrix}}_{\displaystyle = \dot{\bm{\eta}}}.
\end{equation}
Here, $\bm{J}_\mathrm{R}$ is the Jacobian matrix of rotation, which relates the three degrees of freedom for rotational velocity~$\dot{\eta}$ to the angular velocity~$\bm{\omega}$ of the quadrotor.
Furthermore, the angular velocity in $\mathcal{B}$ is given by
\begin{equation}
    \bm{\omega}_\mathcal{B} = \bm{R}^{-1}_{\mathcal{I}\mathcal{B}}(\bm{\eta}) \, \bm{J}_\mathrm{R} \, \dot{\bm{\eta}} = \begin{bmatrix}
        1 & 0 & -\sin{\theta} \\
        0 & \cos{\phi} & \sin{\phi}\cos{\theta} \\
        0 & -\sin{\phi} & \cos{\phi}\cos{\theta}
    \end{bmatrix}\dot{\bm{\eta}},
\end{equation}
where $\bm{R}^{-1}_{\mathcal{I}\mathcal{B}}(\bm{\eta}) = \bm{R}\tran_{\mathcal{I}\mathcal{B}}(\bm{\eta})$.

\subsection{Attitude controller of the Crazyflie}
\begin{figure}[ht]
    \centering
    \includegraphics[scale=0.94]{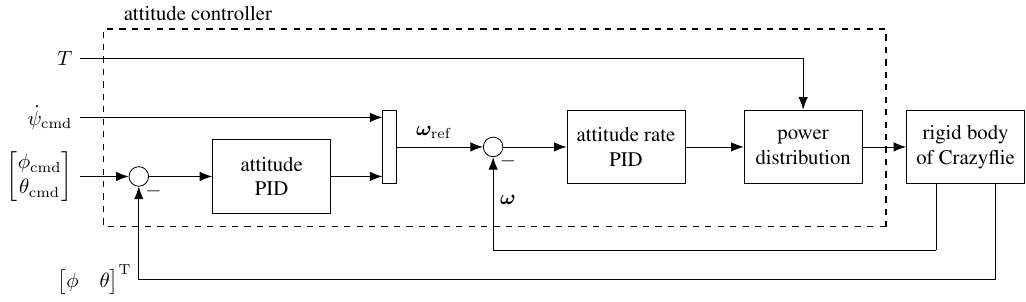}
    \caption{Low-level controlled system of the Crazyflie quadrotor.}
    \label{fig:InnerControlLoop}
\end{figure}

It is necessary to introduce the attitude controller and the resulting low-level controlled system of the Crazyflie quadrotor to derive the equations of motion.
By default, the firmware deployed on the Crazyflie utilizes \acrfull*{pid} controllers to achieve a stable flight behaviour.
The aforementioned inner control loop, illustrated in \Cref{fig:InnerControlLoop}, is tasked with stabilizing the attitude dynamics of the quadrotor and operates at a frequency of $\SI{500}{\hertz}$.
It consists of two \acrshort*{pid} controllers, the \textit{attitude \acrshort*{pid}} and the \textit{attitude rate \acrshort*{pid}}.
The \textit{attitude \acrshort*{pid}} receives a desired roll angle~$\phi_\mathrm{cmd}$ and pitch angle $\theta_\mathrm{cmd}$.
The output of the \textit{attitude \acrshort*{pid}} together with the desired yaw rate $\dot{\psi}_\mathrm{cmd}$ are forwarded in terms of $\bm{\omega}_\mathrm{ref}$ to the \textit{attitude rate \acrshort*{pid}}.
Finally, the \textit{attitude rate \acrshort*{pid}} computes the control signal for the \textit{power distribution}, which together with a desired thrust $T$ calculates the necessary \acrshort{pwm} signals for the rotors.
As mentioned before, a sophisticated \acrshort{mpc} scheme is utilized as guidance law in the outer control loop of \Cref{fig:CompleteControlLoop}.
Therefore, a dynamical model of the low-level controlled system is required.
Based on the derived kinematics and the attitude controller, the equations of motion are derived in the following.

\subsection{Dynamics}
The translational dynamics of the Crazyflie can be described by the Newton equations 
\begin{equation}\label{eq:NewtonEqns}
    \ddot{\bm{\xi}} = -\begin{bmatrix}0 \\ 0 \\ g\end{bmatrix} + \frac{1}{m}\bm{R}_{\mathcal{I}\mathcal{B}}(\bm{\eta})\begin{bmatrix}0 \\ 0 \\ \Delta T+mg\end{bmatrix}
\end{equation}
with the mass~$m$ and the gravitational acceleration~$g$.
The thrust $T = \Delta T + mg$ is separated into a feed-forward part~$mg$ and a differential thrust part~$\Delta T$.
This ensures that for $\bm{\eta}=\bm{0}$ and $\Delta T = 0$, the quadrotor remains in a steady state.
Assuming that the aforementioned low-level control system is running at a sufficiently high frequency, the closed-loop attitude dynamics can be approximated by three first order differential equations
\begin{equation}\label{eq:underlyingDynamics}
    \dot{\phi} = \frac{1}{T_\phi}(\phi_\mathrm{cmd} - \phi), \quad
    \dot{\theta} = \frac{1}{T_\theta}(\theta_\mathrm{cmd} - \theta), \quad \mathrm{and} \quad
    \dot{\psi} = \dot{\psi}_\mathrm{cmd},
\end{equation}
similar to the model introduced in \cite{LlanesKakishWilliamsEtAl24,HuangBauerPan22,KamelStastnyAlexisEtAl17}.
The combined equations of motion~\eqref{eq:NewtonEqns} and \eqref{eq:underlyingDynamics} can then be written in terms of the state-space model
\begin{equation}\label{eq:ApproximatedSystem}
	\dot{\bm{x}} = \bm{f}(\bm{x},\bm{u}) = \begin{bmatrix}
            \bm{\dot{\xi}} \\[0.5em]
            -\begin{bmatrix}0 \\ 0 \\ g\end{bmatrix} + \frac{1}{m}\bm{R}_{\mathcal{I}\mathcal{B}}(\bm{\eta})\begin{bmatrix}0 \\ 0 \\ \Delta T + mg\end{bmatrix} \\[0.5em]
    \frac{1}{T_\phi}(\phi_\mathrm{cmd} - \phi) \\[0.5em]
    \frac{1}{T_\theta}(\theta_\mathrm{cmd} - \theta) \\[0.5em]
    \dot{\psi}_\mathrm{cmd}   
\end{bmatrix}
\end{equation}
with states $\bm{x} = \begin{bmatrix}\bm{\xi} & \dot{\bm{\xi}} & \bm{\eta}\end{bmatrix}\tran \in \mathbb{R}^9$ and inputs $\bm{u} = \begin{bmatrix}\Delta T & \phi_\mathrm{cmd} & \theta_\mathrm{cmd} & \dot{\psi}_\mathrm{cmd}\end{bmatrix}\tran \in \mathbb{R}^4$.
Additionally, the output of the system is defined as $\bm{y} = \bm{h}(\bm{x}) = \begin{bmatrix}x & y & z & \psi\end{bmatrix}\tran \in \mathcal{Y} \subseteq \mathbb{R}^4$ with output-space $\mathcal{Y}$.
Notably, the employed model~\eqref{eq:ApproximatedSystem} comprises the equations of motion for the translation, but only the kinematic description of the rotation due to the employed low-level attitude controller, see Figure~\ref{fig:InnerControlLoop}.

\section{Path-Following Control}\label{sec:PathFollowing}
After introducing the equations of motion for the low-level controlled quadrotor, the path-following problem is defined.
First, the definition of a geometric path is presented.
Furthermore, a short introduction to \acrlong{mpc} is given, followed by the presentation of the \acrfull*{mppfc} algorithm.
\subsection{Geometric Path}
\begin{figure}[ht]
    \centering
    \includegraphics{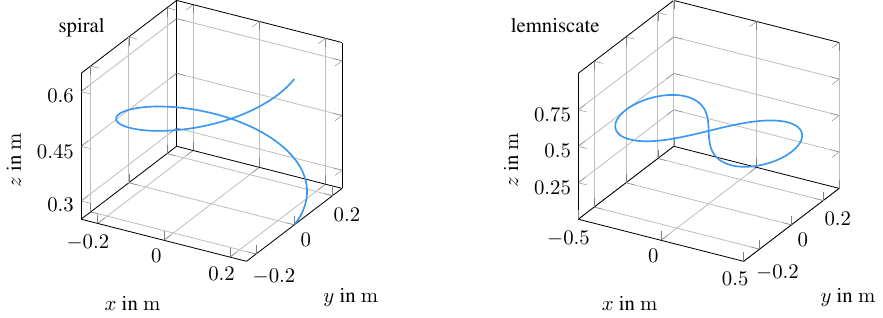}
    \caption{The two paths used to evaluate the performance of the path-following algorithm.}
    \label{fig:PathOverview}
\end{figure}

Throughout this work, the term \textit{path} refers to a geometric curve $\mathcal{P}$ parameterized via a function $\bm{p}(s)\in\mathbb{R}^p$, such that
\begin{equation}\label{eq:PathDefinition}
    \mathcal{P} = \left\{\bm{p}(s) \in \mathcal{Y}\,\vert\,s \in [s_0, 0]\right\},
\end{equation}
where $s \in \mathbb{R}_{\leq 0}$ describes the time-independent path parameter and $s_0 < 0$ its initial condition.
The path is defined in the output-space $\mathcal{Y}$ of the quadrotor.
Choosing the right initial condition for the path parameter is an important design choice of the path-following scheme, as it influences the convergence time to the end of the path.
Without loss of generality, the initial condition of the path parameter is restricted to $s_0 = -1$ in this work.
The performance of the \acrshort*{mpc} based path-following control approach is evaluated based on two different paths, a spiral and a lemniscate with parametric functions
\begin{equation}\label{eq:EvaluationPaths}
    \bm{p}_{\mathrm{spiral}}(s)=\begin{bmatrix}
        0.25\cos{(2\pi s)} \\
        0.25\sin{(2\pi s)} \\
        0.65 + 0.4s \\
        0
    \end{bmatrix}\quad\mathrm{and}\quad
    \renewcommand{\arraystretch}{1.25}
    \bm{p}_{\infty}(s)=\begin{bmatrix}
        0.5\frac{\cos{(2\pi s)}}{\sin^2{(2\pi s)} + 1} \\
        0.5\frac{\sin{(2\pi s)}\cos{(2\pi s)}}{\sin^2{(2\pi s)} + 1} \\
        0.5 \\
        0
    \end{bmatrix},
    \renewcommand{\arraystretch}{1}
\end{equation}
respectively, where the quadrotor shall have zero yaw angle along the path.
Both paths are visualized in \Cref{fig:PathOverview}.

\subsection{Basics of Model Predictive Control}
As illustrated in \Cref{fig:CompleteControlLoop}, an \acrshort{mpc}-based control scheme is utilized in this work.
\acrlong{mpc} is an advanced control concept that operates by repeatedly solving an \acrfull*{ocp}. 
For the Crazyflie quadrotor, described by the differential equation $\dot{\bm{x}} = \bm{f}(\bm{x},\bm{u}) \in \mathbb{R}^n$, at time $t$ the evolution of the states over the prediction horizon $[t, t + T_\mathrm{p}]$ is optimized with respect to a cost functional $J \in \mathbb{R}_{\geq 0}$.
The prediction horizon, defined as the interval of length $T_\mathrm{p}$, enables the \acrshort{mpc} to account for future states of the system.
The control input $\bm{u}  \in \mathbb{R}^m$ of the system is the decision variable to be determined in this optimization problem.
A major conceptual benefit is the explicit inclusion of state and input constraints in the control algorithm, coupled with its capability to effectively manage systems with nonlinear dynamics and multiple inputs and outputs.
A standard form of the \acrshort*{ocp} used in \acrshort*{mpc} is
\begin{subequations}\label{eq:OCP}
\begin{align}
    \underset{\bar{\bm{u}}(\cdot\,\vert\, t)}{\mathrm{minimize}} \quad &J(\bar{\bm{x}}(\cdot\,\vert\, t), \bar{\bm{u}}(\cdot\,\vert\, t)) \label{eq:OCP:CostFunctional} \\
    \mathrm{subject~to} \quad &\dot{\bar{\bm{x}}}(\tau\,\vert\, t) = \bm{f}(\bar{\bm{x}}(\tau\,\vert\, t), \bar{\bm{u}}(\tau\,\vert\, t)), &&\forall \tau \in [t,t+T_\mathrm{p}], \label{eq:OCP:DynamicConstraint} \\ 
    &\bar{\bm{x}}(\tau\,\vert\, t) \in \mathcal{X}, &&\forall \tau \in [t,t+T_\mathrm{p}], \label{eq:OCP:StateConstraint} \\
    &\bar{\bm{u}}(\tau\,\vert\, t) \in \mathcal{U}, &&\forall \tau \in [t,t+T_\mathrm{p}], \label{eq:OCP:InputConstraint} \\
    &\bar{\bm{x}}(t\,\vert\, t) = \bm{x}(t) \label{eq:OCP:InitialCondition}, 
\end{align}
\end{subequations}
where \Cref{eq:OCP:DynamicConstraint} incorporates the system dynamics presented in \Cref{eq:ApproximatedSystem}
and \Cref{eq:OCP:StateConstraint,eq:OCP:InputConstraint} are the state and input constraints of the Crazyflie Quadrotor\cite{RawlingsMayneDiehl17} corresponding to the states and inputs introduced in \Cref{sec:ModellingAndControl}.
A bar over a variable indicates that the variable is to be predicted.
The notation $(\tau\,\vert\, t)$ highlights the existence of two distinct time concepts. 
The first argument represents the time $\tau$ within the time horizon $\tau \in [t, t + T_\mathrm{p}]$, 
while the second argument denotes the time of the experiment $t \in [t_0, t_\mathrm{end}]$ the \acrshort{ocp} is solved at.
Thus, $\bm{\bar{x}}(\tau\,\vert\, t)$ represents the state in the optimization problem, computed based on measurements taken at time $t$ and evaluated at time $\tau$ within the horizon.
The optimal solution to the \acrshort{ocp} is indicated by $\bar{\bm{u}}^*$, which minimizes the positive definite cost functional $J(\bar{\bm{x}}(\cdot\,\vert\, t), \bar{\bm{u}}(\cdot\,\vert\, t))$.
Applying~$\bar{\bm{u}}^*$ to the plant model yields the optimal state trajectory $\bm{\bar{x}}^*$.
The multiple shooting method \cite{RawlingsMayneDiehl17} is employed to solve the \acrshort{ocp} \eqref{eq:OCP} numerically.
Therefore, the prediction horizon is discretized into equidistant time intervals $[t_k, t_k + \delta)$ of length $\delta$, where
\begin{equation}\label{eq:HorizonDiscretization}
    t_k = t + k\delta, \quad k \in \{0, \dots, N-1\},
\end{equation}
and $N = T_\mathrm{p}/\delta$.
Additionally, the input is assumed to have a constant value for time~$\tau \in [t_k,t_k+\delta), \forall k \in \{0, \dots, N-1\}$. 
It should be noted that $t_0 = t$ and $t_{N-1} = t + T_\mathrm{p} - \delta$.
The current state of the plant must be provided as the initial condition in \Cref{eq:OCP:InitialCondition}.
The \acrlong{mpc} algorithm is based on repeatedly solving the \acrshort{ocp} \eqref{eq:OCP}.
After providing the current state of the plant and solving the \acrshort*{ocp}, the control input corresponding to the first time interval of $\bar{\bm{u}}^*$ is applied to the Quadrotor, which is referred to as 
\begin{equation}
    \bm{u}_\mathrm{mpc}(t) = \bar{\bm{u}}^*(\tau\,\vert\,t), \quad \tau \in [t, t+\delta).
\end{equation}
In the next step, the time is incremented by the constant time discretization $\delta$ such that $t \gets t + \delta$ and the algorithm proceeds by measuring the states and solving the \acrshort{ocp} again.
The complete algorithm is as follows
\begin{enumerate}[itemsep=0mm]
    \item Measure the states of the Crazyflie Quadrotor and solve the \acrshort{ocp} (\ref{eq:OCP}).
    \item Apply $\bm{u}_\mathrm{mpc} = \begin{bmatrix}\Delta T & \phi_\mathrm{cmd} & \theta_\mathrm{cmd} & \dot{\psi}_\mathrm{cmd}\end{bmatrix}$ to the plant.
    \item Set $t \gets t + \delta$ and go to step $1$.
\end{enumerate}
Step 2 includes sending the computed control signals to the attitude controller of the Crazyflie quadrotor.

\subsection{Model Predictive Path-Following Control}
The \acrlong{mppfc} algorithm, introduced in~\cite{Faulwasser12}, is a control scheme based on the presented \acrshort{mpc} algorithm in \Cref{eq:OCP} designed to enable accurate tracking of a predefined geometric path as described in \Cref{eq:PathDefinition}.
This approach relies on introducing a timing law that renders the previously time-independent path parameter $s$ time-dependent by means of a virtual input $\nu(t) \in \mathbb{R}$.
This is achieved by influencing the time derivatives of $s$ and, therefore, inducing time in the path parameter.
The virtual input $\nu(t)$ is controlled by the \acrshort{mppfc} algorithm.
Additionally, the timing law is designed to ensure that $\dot{s} > 0$ holds such that the path parameter evolves strictly forward along the path.
This can be easily enforced by defining corresponding state constraints for the path parameter and its derivatives which are included in the MPPFC.
For reasons of simplicity, the timing law is chosen here as an integrator chain of length two, i.e.,
\begin{equation}\label{eq:TimingLaw}
    \underbrace{\begin{bmatrix}
        \dot{s} \\ \ddot{s}
    \end{bmatrix}}_{\displaystyle \dot{\bm{z}}} = \begin{bmatrix}
        0 & 1 \\
        0 & 0 
    \end{bmatrix}
    \underbrace{\begin{bmatrix}
        s \\ \dot{s}
    \end{bmatrix}}_{\displaystyle \bm{z}} + \begin{bmatrix}
        0 \\ 1
    \end{bmatrix}\nu \eqqcolon \bm{g}(\bm{z},\nu)
\end{equation}
such that the virtual input acts as acceleration for the path parameter.
This is specifically chosen such that the time evolution of the path parameter $s$ is continuously differentiable.
The path parameter $s$ and its time derivative $\dot{s}$ are compactly written in the state vector $\bm{z} = \begin{bmatrix}s & \dot{s}\end{bmatrix}\tran$ and are in the following referred to as the states of the path dynamics in \Cref{eq:TimingLaw}.
The path dynamics are subject to state constraints $\mathcal{Z} = [-1, 0] \times (0, \dot{s}_{\mathrm{max}}]$, with an upper bound $\dot{s}_{\mathrm{max}}$ for $\dot{s}$.
Evaluating the path $\bm{p}$ at the now time-dependent path parameter $s(t)$ yields a trajectory which is used as reference for the controller.
The deviation between reference trajectory and the system's output is defined as the path error
\begin{equation}
    \bm{e} = \bm{y} - \bm{p}(s).
\end{equation}
Based on \cite{FaulwasserFindeisen16}, the path error is penalized in \Cref{eq:OCP:CostFunctional} to enforce tracking of the reference trajectory.
Additionally, the path parameter $s$ is penalized to ensure that the path parameter converges to zero, which equals the end of the path.
Furthermore, the input $\bm{u}$ and the virtual input $\nu$ are penalized.
Lastly, to prevent overshoot and aggressive flight manoeuvres, the quadrotor's translational velocity $\dot{\bm{\xi}}$ is penalized.
Consequently, the \acrlong{ocp} for \acrlong{mppfc} is stated as 
\begin{subequations}\label{eq:MPPFC}
    \begin{align}
        \underset{\bar{\bm{u}}(\cdot\,\vert\, t),\, \bar{\nu}(\cdot\,\vert\, t)}{\mathrm{minimize}} \quad &\int_t^{t+T_\mathrm{p}} \left\lVert\begin{matrix}\bar{\bm{e}} \\ \dot{\bar{\bm{\xi}}} \\ \bar{z}_1\end{matrix}\right\rVert_{\bm{Q}}^{2} + \left\lVert\begin{matrix} \bar{\bm{u}} \\ \bar{\nu} \end{matrix}\right\rVert_{\bm{R}}^{2} \,\mathrm{d}\tau + V_\mathrm{f}(\bar{z}_1(t+T_\mathrm{p}\,\vert\, t)) \label{eq:MPPFC:CostFunctional}
    \end{align}
    \vspace{-6mm}
    \begin{align}
        \mathrm{subject~to} \quad &\dot{\bar{\bm{x}}}(\tau\,\vert\, t) = \bm{f}(\bar{\bm{x}}(\tau\,\vert\, t), \bar{\bm{u}}(\tau\,\vert\, t)), &&\forall \tau \in [t,t+T_\mathrm{p}], \label{eq:MPPFC:DynamicConstraintX} \\ 
        &\bar{\bm{x}}(\tau\,\vert\, t) \in \mathcal{X}, \quad&&\forall \tau \in [t,t+T_\mathrm{p}], \label{eq:MPPFC:StateConstraintX} \\
        &\bar{\bm{u}}(\tau\,\vert\, t) \in \mathcal{U}, &&\forall \tau \in [t,t+T_\mathrm{p}], \label{eq:MPPFC:InputConstraintU} \\
        &\bar{\bm{x}}(t\,\vert\, t) = \bm{x}(t), \label{eq:MPPFC:InitialConditionX} \\
        &\bar{\bm{e}}(t+T_\mathrm{p}\,\vert\, t) \in \mathcal{E}_\mathrm{f}, \label{eq:MPPFC:TerminalConditionX} \\
        &\dot{\bar{\bm{z}}}(\tau\,\vert\, t) = \bm{g}(\bar{\bm{z}}(\tau\,\vert\, t), \bar{\nu}(\tau\,\vert\, t)), &&\forall \tau \in [t,t+T_\mathrm{p}], \label{eq:MPPFC:DynamicConstraintZ}\\
        &\bar{\bm{z}}(\tau\,\vert\, t) \in \mathcal{Z}, &&\forall \tau \in [t,t+T_\mathrm{p}], \label{eq:MPPFC:StateConstraintZ} \\
        &\bar{\nu}(\tau\,\vert\, t) \in \mathcal{V}, &&\forall \tau \in [t,t+T_\mathrm{p}], \label{eq:MPPFC:InputConstraintV} \\
        &\bar{\bm{z}}(t\,\vert\, t) = \bm{z}(t), \label{eq:MPPFC:InitialConditionZ} \\
        &\bar{\bm{z}}(t+T_\mathrm{p}\,\vert\, t) \in \mathcal{Z}_\mathrm{f} \label{eq:MPPFC:TerminalConditionZ}
    \end{align}
\end{subequations}
with positive definite matrices $\bm{Q} \in \mathbb{R}^{8\times 8}$ and $\bm{R} \in \mathbb{R}^{5\times 5}$ and $z_i$ denoting the $i$-th entry of the vector $\bm{z}$.
The terminal cost $V_\mathrm{f}$ as well as the terminal constraints $\mathcal{E}_\mathrm{f}$ and $\mathcal{Z}_\mathrm{f}$ can be used to provide nominal stability guarantees.
The computation of stabilizing terminal ingredients lies beyond the scope of this work, as the primary focus is on the practical implementation of the proposed scheme on real hardware.
For further information about the design of terminal ingredients, the reader is referred to \cite{FaulwasserFindeisen16}.

\section{Experimental Results}\label{sec:ExperimentalResults}
The following section evaluates the performance of the proposed path-following control scheme through real-world experiments. To the best of the author's knowledge, this specific approach has not yet been proposed in the literature.
As mentioned in \Cref{sec:Introduction}, the proposed \acrshort{mppfc} scheme is especially well suited for real-world applications, since it not only complies with the state and input constraints of the Crazyflie, but exhibits rapid transient performance and additionally provides a straightforward user-designed tuning oppertunity.

First the experimental setup is described, followed by experimental results for the paths introduced in \Cref{eq:EvaluationPaths}.
Next, a third scenario is presented, which underlines the advantageous properties of the proposed \acrshort{mpc} scheme with respect to state and input constraints in comparison to conventional, two-staged path-following algorithms for quadrotors.
Lastly, the \acrshort{mpc} scheme is extended to a corridor path-following scheme, to compensate possible disadvantages introduced by state and input constraints and providing the controller another degree of freedom for balancing between path advancement and precise output tracking.

\subsection{Experimental Setup}
To minimize disturbances from wind and other environmental factors, and to enable the use of a motion capture system for accurate measurement of position and attitude, all experiments in this section are conducted in an indoor environment.
The Crazyflie 2.0 from Bitcraze \cite{GiernackiSkwierczynskiWitwickiEtAl17} is employed as a quadrotor to evaluate the introduced \acrshort{mpc} scheme for path-following control of a quadrotor.
To ensure safety, a containment structure with a net is utilized, preventing the quadrotor from escaping in the event of a malfunction and hurting humans.
For precise tracking of the Crazyflie's position and attitude, the OptiTrack motion capture system equipped with up to five PrimeX 13 cameras around the workspace is employed.
This system tracks reflective markers on the Crazyflie at a sampling rate of $\SI{240}{\hertz}$, from which the quadrotor's global position and orientation are calculated.
In \Cref{fig:CrazyflieCoordinateSystem} the marker placements on the Crazyflie is illustrated.
The quadrotor's velocity is estimated by applying a finite difference calculation between two consecutive position measurements with a moving average over five consecutive points for smoother velocity estimation.
The attentive reader might have noticed that the paths defined in \Cref{eq:EvaluationPaths} are not starting at ground level.
This is due to the ground effect, which is a difficult to predict phenomenon in quadrotor experiments.
Especially nano quadrotors like the Crazyflie are significantly affected by the ground effect \cite{KanThomasTengEtAl19}.
Therefore, the paths are defined to avoid proximity to the ground and instead the onboard \acrshort{pid} controllers guide the Crazyflie to beginning of the path at $\bm{p}(s_0 = -1)$.
After reaching the start point, the \acrshort{mppfc} takes over control of the Crazyflie.
The OCP \Cref{eq:OCP} is solved by applying the multiple shooting method \cite{RawlingsMayneDiehl17}.
Throughout all the experiments, a prediction horizon of $N=5$ and a time discretization of $\delta = \SI{0.05}{\second}$ is used.
The resulting nonlinear program is formulated using the \verb|CasADi| library \cite{AnderssonGillisHornEtAl19} and consequently solved with the \verb|FATROP| solver \cite{VanroyeSathyaDeSchutterEtAl23}, an interior-point method that utilizes the sparse structure of constraints in \acrshort{ocp}s by employing a specialized linear solver.
The OCP is solved on an external machine and the resulting control signals are transmitted to the Crazyflie via a radio module utilizing the Robot Operating System (ROS) 2 \cite{MacenskiFooteGerkeyEtAl22}.

\subsection{Spiral and Lemniscate}
\begin{figure}[ht]
    \centering
    \includegraphics[scale=0.94]{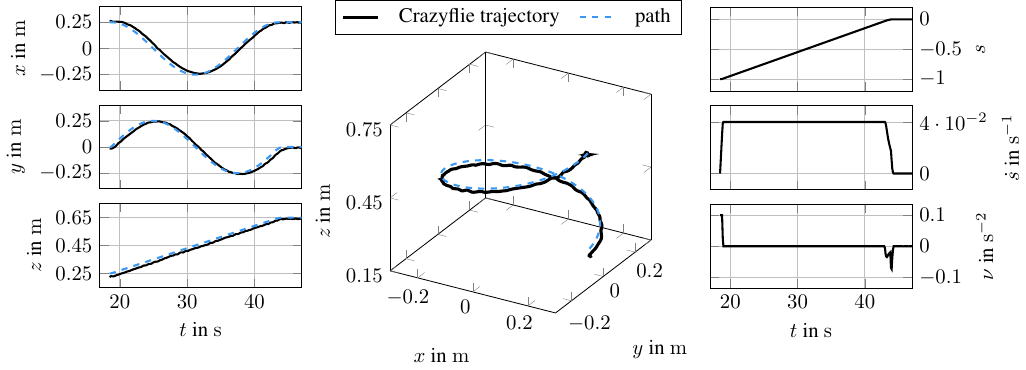}
    \caption{Experimental results of the spiral path-following experiment. The Crazyflie trajectory is shown in black, while the path is depicted in blue.}
    \label{fig:SpiralPathResult}
\end{figure}

In a first experiment, the performance of the presented path-following control scheme is evaluated on the spiral path defined in \Cref{eq:EvaluationPaths}.
The experimental results are illustrated in \Cref{fig:SpiralPathResult}.
Here and in all following equivalent figures, the first column shows the $x$, $y$, and $z$ coordinates of the Crazyflie trajectory as black solid line, and the reference path as dashed blue line.
Overall, a satisfying tracking performance is observed with the quadrotor closely following the path.
A slight deviation in $z$ direction is observed, which causes the quadrotor to constantly operate below the desired height.
This might be caused by slight model mismatches in the thrust generation of the Crazyflie.
Since no integral action or disturbance estimation is included in the proposed \acrshort{mpc} scheme, such a model mismatch induced offset cannot be compensated.
The second (middle) column shows a three-dimensional illustration of the Crazyflie trajectory and the spiral path.
This underlines the results from the first column, where the quadrotor is able to follow the spiral closely, but with a small offset in $z$ direction.
The third column shows the path parameter~$s$ and its derivative~$\dot{s}$ as well as the virtual input~$\nu$.
The virtual input accelerates $\dot{s}$ to the state constraint $\dot{s}_\mathrm{max} = \SI{0.04}{\per\second}$ and then keeps it constant until the end of the path is reached.
This leads to linear convergence of the path parameter $s$ to the origin.
It is observed in the experiments that the tracking performance for the orientation is as equally good as for the position.

\begin{figure}[ht]
    \centering
    \includegraphics[scale=0.94]{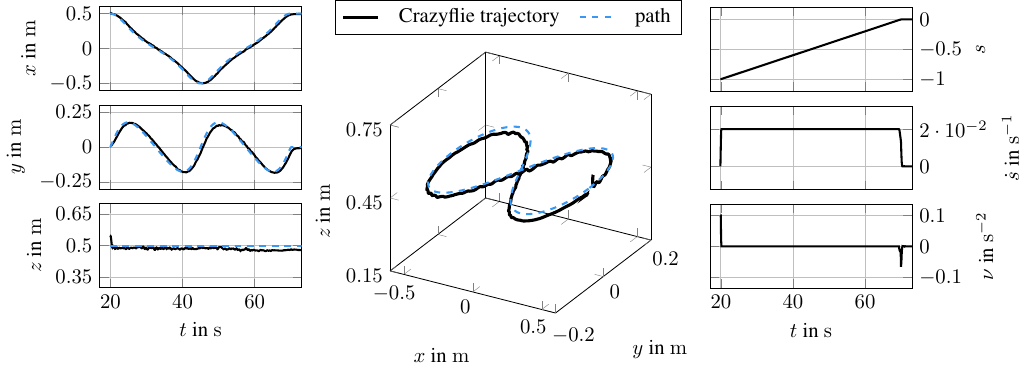}
    \caption{Experimental results of the lemniscate path-following experiment. The Crazyflie trajectory is shown in black, while the path is depicted in blue.}
    \label{fig:LemniscatePathResult}
\end{figure}

\begin{figure}[ht]
    \centering
    \includegraphics[scale=0.08]{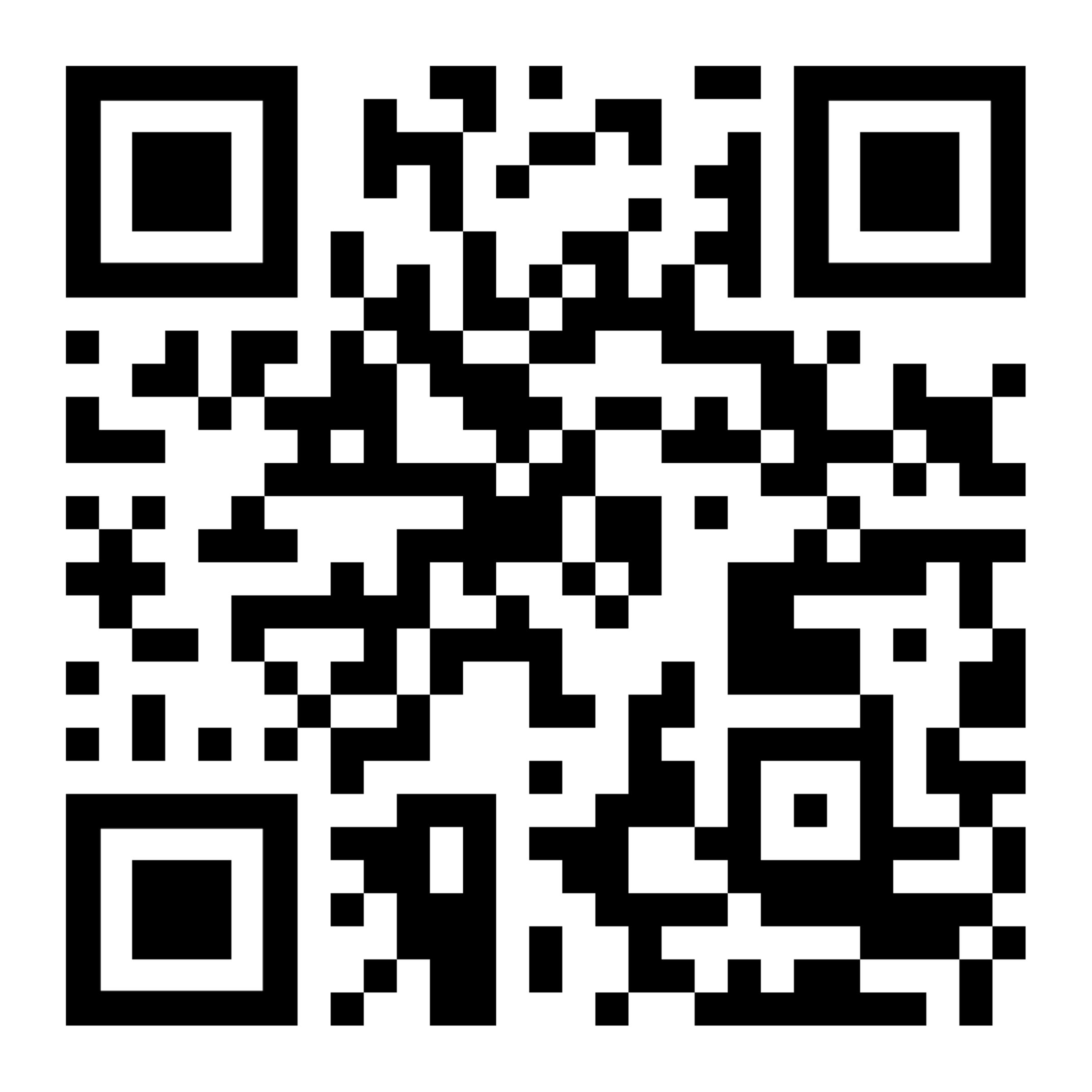}
    \includegraphics[scale=0.07]{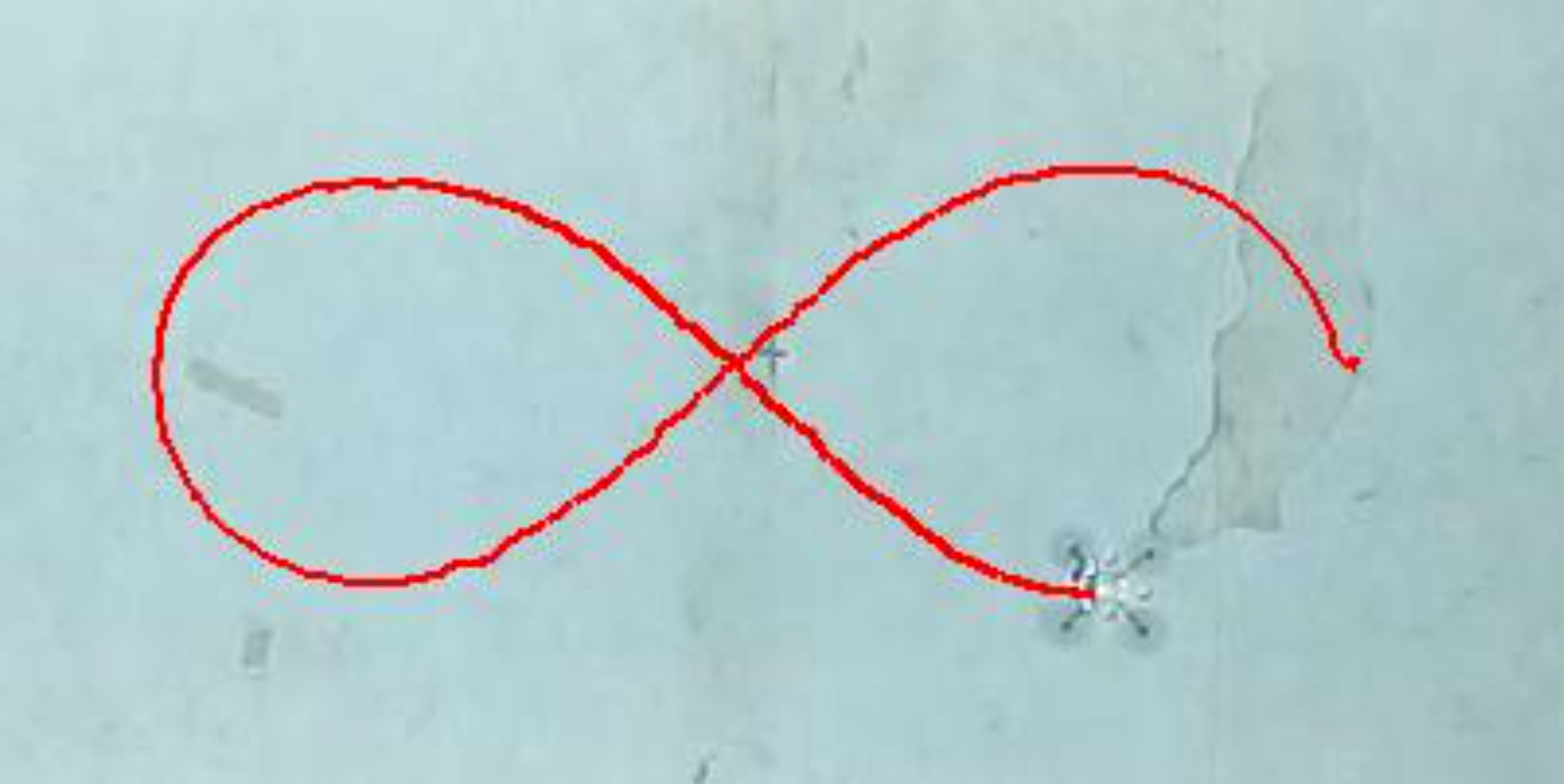}
    \caption{Video of the lemniscate path-following experiment. The QR code links to the video online.}
    \label{fig:ExperimentVideo}
\end{figure}

In a second experiment, the performance of the presented path-following control scheme is evaluated on the lemniscate, as defined in \Cref{eq:EvaluationPaths}.
The results are depicted in \Cref{fig:LemniscatePathResult}.
Similar to \Cref{fig:SpiralPathResult}, the first column shows the $x$, $y$, and $z$ coordinates of the Crazyflie trajectory as black solid line, and the reference path as dashed blue line.
Analogously to the spiral path, the quadrotor emits sufficient tracking of the reference, with a qualitatively similar offset in $z$ direction.
The second (middle) column shows the three-dimensional representation of the Crazyflie trajectory and the lemniscate path, which underlines the results from the first column.
Identically to the spiral path experiment, the velocity of the path parameter operates at its maximum throughout the whole experiment, leading to linear convergence of the path parameter to the origin, as displayed in the third column.
A video of the lemniscate path-following experiment is available online by scanning the QR code in \Cref{fig:ExperimentVideo}.

\subsection{Advantages of MPPFC}
\begin{figure}[ht]
    \centering
    \includegraphics{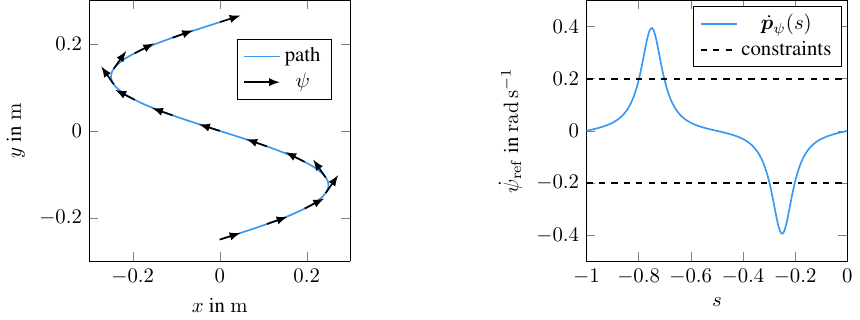}
    \caption{Sinusoidal path with orientation aligned tangentially to the curvature.}
    \label{fig:Sinus2DOrientation}
\end{figure}

\begin{figure}[ht]
    \centering
    \includegraphics{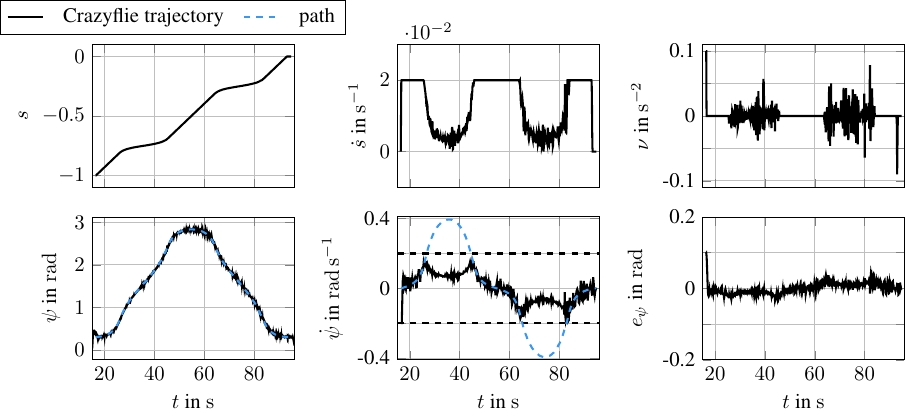}
    \caption{Path parameter $s$ and its derivative $\dot{s}$ for the sinusoidal path.}
    \label{fig:SinusPathResultsYaw}
\end{figure}

\begin{figure}[ht]
    \centering
    \includegraphics[scale=0.96]{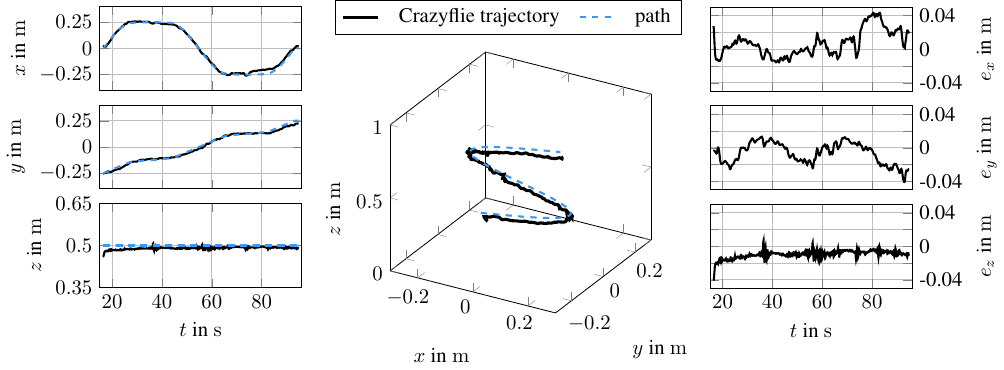}
    \caption{Experimental results of the sinusoidal path-following experiment. The Crazyflie trajectory is shown in black, while the path is depicted in blue.}
    \label{fig:SinusPathResults}
\end{figure}

So far, the presented results include only cases where the path parameter converges linearly to the origin.
Consequently, the quadrotor itself does not operate at its limits, as neither the quadrotor's state nor input constraints are active, which would be one of the major advantages of MPC.
Instead, the constraint on the path parameter's derivative determines the actual path following behaviour.
A similar result is achievable with the previously mentioned path-following controllers from other works.
The following third real-world scenario exemplifies the major advantage of the one-stage model-based path-following control scheme.
The Crazyflie shall track a sinusoidal path
\begin{equation}\label{eq:SinusPath}
    \bm{p}(s) = \begin{bmatrix}
        0.25\sin{(2\pi s)} & 0.25 + 0.5s & 0.5 & \mathrm{arctan2}(0.5,0.5\pi\cos{(2\pi s)})
    \end{bmatrix}^\mathrm{T},
\end{equation} 
where $\mathrm{arctan2}(y,x)$ denotes the four-quadrant inverse tangent function.
The orientation $\bm{p}_\psi(s)$ of \Cref{eq:SinusPath} defines the yaw angle of the quadrotor to be always tangential to the path, which is illustrated in the left plot of \Cref{fig:Sinus2DOrientation}.
This is a common requirement in practice, for example, when the quadrotor is transporting a payload or has sensors mounted on it.
In particular, the quadrotor's yaw rate shall here be constrained to $\dot{\psi} \in [\SI{-0.2}{\radian\per\second}, \SI{0.2}{\radian\per\second}]$.
Assuming that the quadrotor is following the path exactly, the yaw rate can be calculated based only on the path $\bm{p}_\psi(s)$ and the path dynamics $\bm{z}=\begin{bmatrix}s &  \dot{s} \end{bmatrix}^\top$ as
\begin{equation}\label{eq:YawRate}
    \dot{\bm{p}}_\psi(s) = \frac{\partial \bm{p}_\psi(s)}{\partial s}\dot{s} = \frac{2\pi^2\sin{(2\pi s)}}{\pi^2\cos{(2\pi s)}^2 + 1}\dot{s}.
\end{equation}
Then, considering the previous scenario, where the path parameter converged linearly to the origin and a maximum path parameter velocity of $\dot{s} = \SI{0.02}{\per\second}$, $\dot{\bm{p}}_\psi(s)$ would exceed the desired yaw rate constraint significantly, as seen in the right plot of \Cref{fig:Sinus2DOrientation}.
The path dynamics and the dynamics of the quadrotor are both part of the \acrshort{ocp}, as a consequence, the \acrshort{mppfc} reduces the path parameter velocity to comply with the yaw rate constraint.
The experimental results are illustrated in \Cref{fig:SinusPathResultsYaw,fig:SinusPathResults}.
In contrast to the previous results in \Cref{fig:SpiralPathResult,fig:LemniscatePathResult}, the path parameter $s$ does not converge linearly to the origin.
Instead, two plateaus at $t = \SI{35}{\second}$ and $t = \SI{70}{\second}$ are observed, as depicted in the top left plot of \Cref{fig:SinusPathResultsYaw}.
These plateaus are caused by a reduced path parameter velocity $\dot{s}$ as illustrated in the top middle plot of \Cref{fig:SinusPathResultsYaw}.
There, $\dot{s}$ is reduced almost to zero to comply with the constraint on the yaw rate, since a reduced convergence speed of the path parameter leads to a reduced yaw rate, as seen in \Cref{eq:YawRate}.
This is underlined in the bottom middle plot of \Cref{fig:SinusPathResultsYaw}, the reduced path parameter velocity occurs when the nominal $\dot{\bm{p}}_\psi$, depicted in a dashed blue line, is about to exceed the constraint, coloured in the dashed black line.
Note, that the \acrshort{mpc} successfully keeps the yaw rate within its defined constraints.
Despite the change in path parameter velocity, the quadrotor is still able to follow the reference yaw angle closely, while complying with the constraints, as shown in the bottom left and bottom right plots in \Cref{fig:SinusPathResultsYaw}.
Additionally, similar to previous results, good tracking performance in $x$, $y$, and $z$ direction is observed, as shown in \Cref{fig:SinusPathResults}.

\subsection{Corridor Path-Following}
\begin{figure}[ht]
    \centering
    \includegraphics{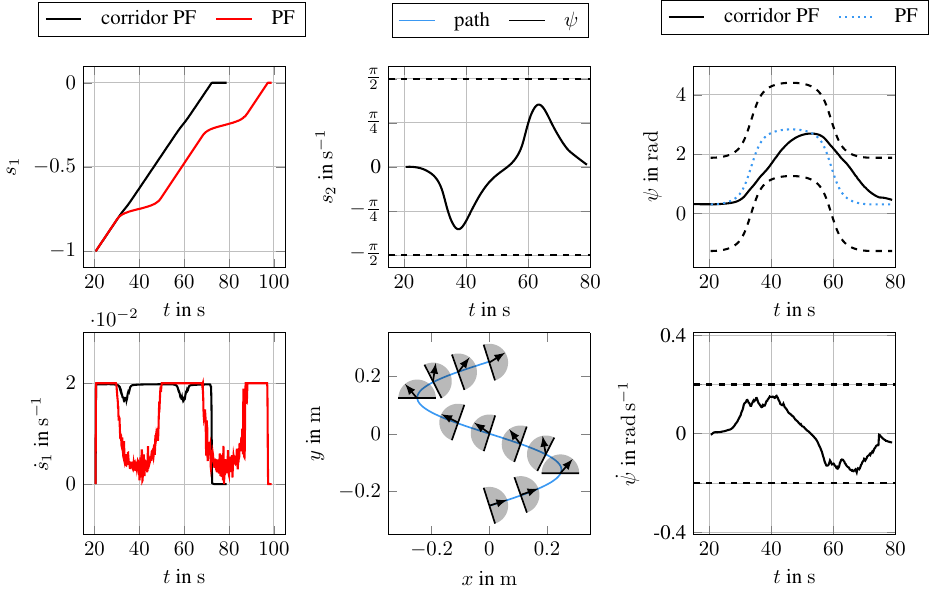}
    \caption{Experimental results for the sinusoidal path with corridor path-following.}
    \label{fig:Sinus2}
\end{figure}

For a quadrotor, strictly following the path can sometimes be too restrictive, for instance, allowing deviations might be beneficial if it leads to cost reductions.
Therefore, the concept of corridor path-following is introduced in the following.
The core idea is to define a corridor around the reference path within which deviations are permitted.
This flexibility might lead to a faster path following and therefore improves costs and shortens the operation time, which is a crucial factor in real-world applications.
Especially in the context of quadrotors, where the energy supplied by the battery is limited, a longer operation time can lead to a reduced range of the quadrotor.
As seen in the results in \Cref{fig:SpiralPathResult,fig:LemniscatePathResult}, in the case where no constraints are active, the upper limit of the path parameter velocity $z_{2,\mathrm{max}}$ is the leading factor for determining the time until the path parameter converged to the origin.
While this constraint can be designed a-priori, possible state and input constraints might affect the path parameter velocity, as seen in the previous subsection.
As a consequence, the operation time of the quadrotor is prolonged.
By granting the quadrotor limited flexibility through a corridor along the path, it becomes easier to satisfy the state constraints and therefore can reduce the operation time. 
These deviations are governed and penalized by an additional path parameter $s_2$, equipped with its own dynamics and a corresponding virtual input $\nu_2$. 
The allowable size of the corridor is enforced through state constraints on $s_2$, ensuring that deviations from the nominal path remain within a user-defined region. 
Therefore, the presented path-following control scheme is used to enable faster convergence of the path parameter to the origin by permitting controlled deviations from the path within a predefined corridor in the output space.
In consistency with the previously introduced example of the sinusoidal path, $s_2$ is used to allow for some predefined maximum deviations in the $\psi$ direction.
The modified path then reads 
\begin{align}
    \bm{p}_\mathrm{c}(s_1,s_2) = \begin{bmatrix}
        0.25\sin(2\pi s_1) \\
        0.25 + 0.5s_1 \\ 
        0.5 \\ 
        \mathrm{arctan2}(0.5,0.5\pi\cos(2\pi s_1))
    \end{bmatrix} + \begin{bmatrix}
        0 \\
        0 \\
        0 \\
        s_2
    \end{bmatrix},
\end{align}
with $s_2 \in [-0.5\pi,0.5\pi]$, which defines the corridor of possible deviations for the orientation. 
Notably, this additional parameter has the origin in its interior which is in contrast to the original path parameter~$s_1$.
The new path dynamics is
\begin{equation}
    \underbrace{\begin{bmatrix}
        \dot{s}_1 \\ 
        \dot{s}_2 \\ 
        \ddot{s}_1 \\
        \ddot{s}_2
    \end{bmatrix}}_{\dot{\bm{z}}(t)} = \begin{bmatrix}
        0 & 0 & 1 & 0\\
        0 & 0 & 0 & 1\\
        0 & 0 & 0 & 0 \\
        0 & 0 & 0 & 0
    \end{bmatrix}
    \underbrace{\begin{bmatrix}
        s_1 \\ 
        s_2 \\
        \dot{s}_1 \\
        \dot{s}_2
    \end{bmatrix}}_{\bm{z}(t)} + \begin{bmatrix}
        0 & 0\\
        0 & 0 \\
        1 & 0 \\
        0 & 1
    \end{bmatrix}\underbrace{\begin{bmatrix}
        \nu_1 \\
        \nu_2
    \end{bmatrix}}_{\bm{\nu}(t)} = \bm{g}(\bm{z}(t),\bm{\nu}(t))
\end{equation}
and the new OCP is
\begin{align}
    \underset{\bar{\bm{u}}(\cdot\,\vert\, t)\ ,\bar{\bm{\nu}}(\cdot\,\vert\, t)}{\mathrm{minimize}} \quad &\int_t^{t+T_\mathrm{p}} \left\lVert\begin{matrix}\bar{\bm{e}} \\ \dot{\bar{\bm{\xi}}} \\ \bar{z}_1 \\\bar{z}_2\end{matrix}\right\rVert_{\bm{Q}}^{2} + \left\lVert\begin{matrix} \bar{\bm{u}} \\ \bar{\bm{\nu}} \end{matrix}\right\rVert_{\bm{R}}^{2} \,\mathrm{d}\tau + V_\mathrm{f}(\bar{z}_1(t+T_\mathrm{p}\,\vert\, t),\bar{z}_2(t+T_\mathrm{p}\,\vert\, t)) \label{eq:MPCPFC:CostFunctional}
\end{align}
with the same constraints as the \acrshort{ocp} in \Cref{eq:MPPFC} and which is solved in the same manner as previously described.
The results of the proposed corridor path-following scheme are shown in \Cref{fig:Sinus2}, where the tracking of the sinusoidal path with and without additional corridor is compared by means of simulations.
By allowing for some deviations in tracking the path in a tangential fashion with respect to the yaw angle~$\psi$, cf.\ the bottom middle plot of \Cref{fig:Sinus2},
the resulting operation time decreases by around $\SI{25}{\second}$, which is a benefit of approximately $\SI{35}{\percent}$ compared to the previous results shown in \Cref{fig:SinusPathResultsYaw}.
This can be observed in the top left plot of \Cref{fig:Sinus2}, where the path parameter~$s_1$ in the case of corridor path-following, coloured in black, is able to follow the path closely, without slowing down the path parameter and thus the quadrotor.
In comparison, the evolution of the path parameter of the original path-following scheme presented in \Cref{fig:SinusPathResultsYaw} is coloured in red.
This behaviour  is underlined by the evolution of the path parameter velocity, shown in the bottom left plot of \Cref{fig:Sinus2}.
While the path parameter velocity~$\dot{s}_1$ still needs to be reduced, the amount and duration of the reduction is significantly less compared to the classical path-following approach.
The reason for this behaviour is that the quadrotor is able to make use of the relaxation path parameter $s_2$ and thus is deviating from the nominal path to avoid actions with high yaw rates.
This can be seen in the bottom middle plot of \Cref{fig:Sinus2}.
There, the orientation $\psi$ of the quadrotor is not aligned tangentially to the path anymore, but instead deviates as mentioned above.
As seen in the top right plot of \Cref{fig:Sinus2}, the deviation in orientation remains only within the defined admissible corridor.
Additionally, this is illustrated by cone shapes in the bottom middle plot of \Cref{fig:Sinus2}, which visualize the admissible corridor for the orientation.
It is observed that the quadrotor makes use of the complete corridor to avoid high yaw rates and as a consequence would reduce the path parameter velocity, since a high path parameter velocity guarantees fast converges to the end of the path and therefore resembles a cost improvement.
Additionally, the deviation vanishes towards the end of the path.

\section{Summary}\label{sec:Conclusion}
This paper presented the application of an \acrlong{mpc}-based path-following control framework to the Crazyflie quadrotor.
By employing a cascaded control structure, the proposed \acrlong{mppfc} served as the guidance law, generating desired attitude and thrust commands for the inner-loop attitude controller.
This architecture enabled real-time implementation and was validated through real-world experiments across three distinct scenarios.
The first two scenarios demonstrated the feasibility and usability of the method as a proof of concept. The third scenario highlighted the advantages of the \acrshort{mpc}-based path-following approach over alternative solutions from the literature.
To address cases where strict path adherence is overly restrictive, a corridor-based path-following extension was proposed. 
This extension effectively balanced path deviations with performance improvements in terms of cost.
Future work may involve extending the method to a swarm of quadrotors. 
The \acrshort{mpc}-based formulation lends itself to a distributed control scheme, enabling decentralized coordination among multiple agents.
Another promising direction is the integration of collision avoidance via path-finding algorithms. 
While this work relied on predefined paths, adapting the method for flexible, dynamic path planning could enable its use in more complex and changing environments.

\section{Acknowledgements}
This work was supported by the Deutsche Forschungsgemeinschaft (DFG, German Research Foundation) under project EB195/40-1, 501890093 ``Mehr Intelligenz wagen - Designassistenten in Mechanik und Dynamik (SPP 2353)'' and under Germany's Excellence Strategy - EXC 2075 - 390740016, project PN4-4 ``Learning from Data - Predictive Control in Adaptive Multi-Agent Scenarios''.


\begin{thebibliography}{10}
\expandafter\ifx\csname url\endcsname\relax
  \def\url#1{\texttt{#1}}\fi
\expandafter\ifx\csname urlprefix\endcsname\relax\def\urlprefix{URL }\fi
\expandafter\ifx\csname href\endcsname\relax
  \def\href#1#2{#2} \def\path#1{#1}\fi

\bibitem{MaslekarKulkarniChakravarthy20}
N.~V. Maslekar, K.~P. Kulkarni, A.~K. Chakravarthy, Application of {{Unmanned
  Aerial Vehicles}} ({{UAVs}}) for {{Pest Surveillance}}, {{Monitoring}} and
  {{Management}}, in: A.~K. Chakravarthy (Ed.), Innovative {{Pest Management
  Approaches}} for the 21st {{Century}}, Springer, Singapore, 2020, pp. 27--45.

\bibitem{SkjetneFossenKokotovic04}
R.~Skjetne, T.~I. Fossen, P.~V. Kokotovi{\'c}, Robust output maneuvering for a
  class of nonlinear systems, Automatica 40~(3) (2004) 373--383.

\bibitem{ChenXuEbelEtAl25}
J.~Chen, L.~Xu, H.~Ebel, P.~Eberhard, An {{Online Optimization-Based Trajectory
  Planning Approach}} for {{Cooperative Landing Tasks}} (Feb. 2025).

\bibitem{RubiMorcegoPerez20}
B.~Rubi, B.~Morcego, R.~Perez, A {{Deep Reinforcement Learning Approach}} for
  {{Path Following}} on a {{Quadrotor}}, in: 2020 {{European Control
  Conference}} ({{ECC}}), Saint Petersburg, Russia, 2020, pp. 1092--1098.

\bibitem{PerezLeonAcevedoMillanRomeraEtAl20}
H.~{Perez-Leon}, J.~J. Acevedo, J.~A. {Millan-Romera}, A.~{Castillejo-Calle},
  I.~Maza, A.~Ollero, An aerial robot path follower based on the `carrot
  chasing' algorithm, in: M.~F. Silva, J.~Lu{\'i}s~Lima, L.~P. Reis,
  A.~Sanfeliu, D.~Tardioli (Eds.), Robot 2019: {{Fourth}} Iberian Robotics
  Conference, Springer International Publishing, Cham, 2020, pp. 37--47.

\bibitem{RubiPerezMorcego20}
B.~Rub{\'i}, R.~P{\'e}rez, B.~Morcego, A {{Survey}} of {{Path Following Control
  Strategies}} for {{UAVs Focused}} on {{Quadrotors}}, Journal of Intelligent
  \& Robotic Systems 98~(2) (2020) 241--265.

\bibitem{RozaMaggiore12}
A.~Roza, M.~Maggiore, Path following controller for a quadrotor helicopter, in:
  {{American Control Conference}} ({{ACC}}), Montreal, 2012, pp. 4655--4660.

\bibitem{Faulwasser12}
T.~Faulwasser, Optimization-based {{Solutions}} to {{Constrained
  Trajectory-tracking}} and {{Path-following Problems}}, Doctoral thesis,
  Otto-von-Guericke-Universit{\"a}t, Magdeburg, Germany (2012).

\bibitem{FaulwasserFindeisen16}
T.~Faulwasser, R.~Findeisen, Nonlinear {{Model Predictive Control}} for
  {{Constrained Output Path Following}}, IEEE Transactions on Automatic Control
  61~(4) (2016) 1026--1039.

\bibitem{FaulwasserWeberZometaEtAl17}
T.~Faulwasser, T.~Weber, P.~Zometa, R.~Findeisen, Implementation of {{Nonlinear
  Model Predictive Path-Following Control}} for an {{Industrial Robot}}, IEEE
  Transactions on Control Systems Technology 25~(4) (2017) 1505--1511.

\bibitem{RomeroSunFoehnEtAl22}
A.~Romero, S.~Sun, P.~Foehn, D.~Scaramuzza, Model predictive contouring control
  for time-optimal quadrotor flight, IEEE Transactions on Robotics 38~(6)
  (2022) 3340--3356.

\bibitem{WangPanHuEtAl19}
D.~Wang, Q.~Pan, J.~Hu, C.~Zhao, Y.~Guo, {{MPCC-based Path Following Control}}
  for a {{Quadrotor}} with {{Collision Avoidance Guaranteed}} in {{Constrained
  Environments}}, in: 28th {{International Symposium}} on {{Industrial
  Electronics}} ({{ISIE}}), Vancouver, 2019, pp. 581--586.

\bibitem{WangZhaoHuEtAl20}
D.~Wang, C.~Zhao, J.~Hu, Q.~Pan, Model {{Predictive Path Following Control}} of
  a {{Quadrotor}} in {{Constrained Environments}}, in: 2020 {{IEEE}} 16th
  {{International Conference}} on {{Control}} \& {{Automation}} ({{ICCA}}),
  Singapore, 2020, pp. 719--724.

\bibitem{SanchezDJorgeRaffoEtAl21}
I.~S{\'a}nchez, A.~D'Jorge, G.~V. Raffo, A.~H. Gonz{\'a}lez, A.~Ferramosca,
  Nonlinear {{Model Predictive Path Following Controller}} with {{Obstacle
  Avoidance}}, Journal of Intelligent \& Robotic Systems 102~(1) (2021) paper 16.

\bibitem{WeiZhengLiEtAl24}
M.~Wei, L.~Zheng, H.~Li, H.~Cheng, Adaptive {{Neural Network-Based Model
  Path-Following Contouring Control}} for {{Quadrotor Under Diversely Uncertain
  Disturbances}}, IEEE Robotics and Automation Letters 9~(4) (2024) 3751--3758.

\bibitem{YangZhengPanEtAl21}
R.~Yang, L.~Zheng, J.~Pan, H.~Cheng, Learning-{{Based Predictive Path Following
  Control}} for {{Nonlinear Systems Under Uncertain Disturbances}}, IEEE
  Robotics and Automation Letters 6~(2) (2021) 2854--2861.

\bibitem{GrosZanonQuirynenEtAl20}
S.~Gros, M.~Zanon, R.~Quirynen, A.~Bemporad, M.~Diehl, From linear to nonlinear
  {{MPC}}: Bridging the gap via the real-time iteration, International Journal
  of Control 93~(1) (2020) 62--80.

\bibitem{LlanesKakishWilliamsEtAl24}
C.~Llanes, Z.~Kakish, K.~Williams, S.~Coogan, {{CrazySim}}: {{A
  Software-in-the-Loop Simulator}} for the {{Crazyflie Nano Quadrotor}}, in:
  {{IEEE International Conference}} on {{Robotics}} and {{Automation}}
  ({{ICRA}}), Yokohama, 2024, pp. 12248--12254.

\bibitem{HuangBauerPan22}
Z.~Huang, R.~Bauer, Y.-J. Pan, Closed-{{Loop Identification}} and {{Real-Time
  Control}} of a {{Micro Quadcopter}}, IEEE Transactions on Industrial
  Electronics 69~(3) (2022) 2855--2863.

\bibitem{KamelStastnyAlexisEtAl17}
M.~Kamel, T.~Stastny, K.~Alexis, R.~Siegwart, Model predictive control for
  trajectory tracking of unmanned aerial vehicles using robot operating system,
  in: A.~Koubaa (Ed.), Robot Operating System ({{ROS}}): {{The}} Complete
  Reference (Volume 2), Springer International Publishing, Cham, 2017, pp.
  3--39.

\bibitem{RawlingsMayneDiehl17}
J.~B. Rawlings, D.~Q. Mayne, M.~Diehl, Model {{Predictive Control}}:
  {{Theory}}, {{Computation}}, and {{Design}}, 2nd Edition, Nob Hill
  Publishing, Madison, 2017.

\bibitem{GiernackiSkwierczynskiWitwickiEtAl17}
W.~Giernacki, M.~Skwierczynski, W.~Witwicki, P.~Wronski, P.~Kozierski,
  Crazyflie 2.0 quadrotor as a platform for research and education in robotics
  and control engineering, in: 22nd {{International Conference}} on {{Methods}}
  and {{Models}} in {{Automation}} and {{Robotics}} ({{MMAR}}), Miedzyzdroje,
  2017, pp. 37--42.

\bibitem{KanThomasTengEtAl19}
X.~Kan, J.~Thomas, H.~Teng, H.~G. Tanner, V.~Kumar, K.~Karydis, Analysis of
  {{Ground Effect}} for {{Small-Scale UAVs}} in {{Forward Flight}}, IEEE
  Robotics and Automation Letters 4~(4) (2019) 3860--3867.

\bibitem{AnderssonGillisHornEtAl19}
J.~A.~E. Andersson, J.~Gillis, G.~Horn, J.~B. Rawlings, M.~Diehl, {{CasADi}}: A
  software framework for nonlinear optimization and optimal control,
  Mathematical Programming Computation 11~(1) (2019) 1--36.

\bibitem{VanroyeSathyaDeSchutterEtAl23}
L.~Vanroye, A.~Sathya, J.~De~Schutter, W.~Decr{\'e}, {{FATROP}}: {{A Fast
  Constrained Optimal Control Problem Solver}} for {{Robot Trajectory
  Optimization}} and {{Control}}, in: {{{{IEEE}}/{{RSJ}} International
  Conference}} on {{Intelligent Robots}} and {{Systems}} ({{IROS}}), Detroit,
  2023, pp. 10036--10043.

\bibitem{MacenskiFooteGerkeyEtAl22}
S.~Macenski, T.~Foote, B.~Gerkey, C.~Lalancette, W.~Woodall, Robot {{Operating
  System}} 2: {{Design}}, architecture, and uses in the wild, Science Robotics
  7~(66) (2022) paper eabm6074.

\end{thebibliography}

\end{document}